\documentstyle[12pt]{article}
\begin{document}  

\title{\bf Gravity from Dirac Eigenvalues
\footnote{\ This essay received an
``honorable mention'' from the Gravity Research Foundation, 1997 --- Ed.}
}
\author{Giovanni Landi${}^1$\footnote{\ 
Fellow of the Italian National Council of Research (CNR)
under Grant CNR-NATO 215.29/01. 
On leave from Dipartimento di Scienze Matematiche,
Universit\`a di Trieste, P.le Europa 1, I-34127, Trieste, Europe.
Also, INFN, Sezione di Napoli, Napoli, Europe.} , 
Carlo Rovelli${}^2$  \\ ~ \\
${}^1$ DAMTP, University of Cambridge, \\
Silver St. Cambridge, CB3 9EW, Europe.
\\ 
landi@univ.trieste.it \\ 
${}^2$ Physics Department, University of Pittsburgh, \\
Pittsburgh Pa 15260, USA. \\
rovelli@pitt.edu}
\maketitle

\begin{abstract}

We study a formulation of euclidean general relativity in which the 
dynamical variables are given by a sequence of real numbers 
$\lambda_{n}$, representing the eigenvalues of the Dirac operator on 
the curved spacetime.  These quantities are diffeomorphism-invariant 
functions of the metric and they form an infinite set of ``physical 
observables'' for general relativity.  Recent work of Connes and 
Chamseddine suggests that they can be taken as natural variables for 
an invariant description of the dynamics of gravity.  We compute the 
Poisson brackets of the $\lambda_{n}$'s, and find that these can be 
expressed in terms of the propagator of the linearized Einstein 
equations and the energy-momentum of the eigenspinors.  We show that 
the eigenspinors' energy-momentum is the Jacobian matrix of the change 
of coordinates from the metric to the $\lambda_{n}$'s.  We study a 
variant of the Connes-Chamseddine spectral action which eliminates a 
disturbing large cosmological term.  
We analyze the corresponding equations of 
motion and find that these are solved if the energy momenta of the 
eigenspinors scale linearly with the mass.  Surprisingly, this scaling 
law codes Einstein's equations.
Finally we study the coupling to a physical fermion field.
\end{abstract}
\bigskip\bigskip

\newpage
\section{Introduction}

In this paper, we present a novel formulation of euclidean general 
relativity.  More precisely, we study a theory that approximates 
general relativity at large scale while modifying it at 
short scale.  The theory is characterized by the fact that it is not 
formulated strictly as a field theory.  The dynamical variables are 
not fields: they are an (infinite) sequence of real numbers 
$\lambda_{n}$, $n$ being an integer.  The relation between 
the usual representation of the gravitational field by means of the metric 
tensor $g_{\mu\nu}(x)$, and the representation in terms of the 
$\lambda_{n}$'s is given by the fact that the $\lambda_{n}$'s are the 
eigenvalues of the Dirac operator $D$ defined on the spacetime 
geometry described by $g_{\mu\nu}(x)$.

Thus, the general idea is to describe spacetime geometry by 
giving the eigen-frequencies of the spinors that can live on that 
spacetime.  This is in the spirit of the well known mathematical problem: 
``Can one hear the shape of a drum?'' \cite{drum}.  Namely the problem of 
characterizing a two-dimensional shape by means of the spectrum of the 
laplacian defined on that shape. (Within some approximation, this spectrum 
gives the Fourier decomposition of the sound emitted by a drum with 
that shape.)

The theory we study is implicitly contained in the recent work of 
Connes and Chamseddine \cite{alain}.  Alain Connes' exciting and 
ambitious attempt to unravel a microscopic noncommutative structure of 
spacetime \cite{alainb,lot,daniel3,landi} has generated, among several 
others, also the idea that the Dirac operator $D$ encodes the full 
information about the spacetime geometry in a way usable for 
describing gravitational dynamics.  First of all the geometry 
can be reconstructed from $D$.  More precisely it can be reconstructed 
from the (normed) algebra generated by $D$ and by the smooth functions 
$f$ on spacetime.  If $x$ and $y$ are spacetime points, then their
geodesic distance can be expressed 
in terms of $D$ \cite{alainb} as  
\begin{equation} 
d(x,y) = sup_{f}\ \{|f(x)-f(y)| ~:~ |\!|[D, f]|\!| \le 1 \}. 
\end{equation}
The $sup$ is over all functions $f$ whose commutator with $D$ has norm 
less than one (here $f$ and $D$ are viewed as operators acting on the 
Hilbert space of the spinors on the manifold, and the norm is the 
natural operator norm).  Secondly, there is a natural way 
of giving the dynamics in terms of $D$: The Einstein-Hilbert 
action with a cosmological term is approximated by the 
trace of a very simple function of $D$ \cite{alain}, as we recall 
below.  These results suggest that one can take the Dirac operator $D$ 
as the object representing the dynamical field, and try to develop a 
dynamical theory for $D$.

In fact, in \cite{alain}, the powerful machinery of non-commutative 
geometry is used to elegantly encode the Yang-Mills structure of the 
standard model into a non-commutative component of the spacetime 
geometry.  Accordingly, (a generalized operator) $D$ codes the 
gravitational field {\it and\ } the Yang-Mills fields, plus, as a very 
remarkable bonus, the Higgs fields as well.  However, even 
independently from these results on the standard model, we think that 
the idea of encoding the dynamical field into $D$ is very interesting 
also for gravity alone.  This is the idea we pursue here.  We take the 
purely gravitational component of the Connes-Chamseddine theory only 
and we remain in the regime of conventional commutative geometry.

More in detail, we consider the idea of encoding the gravitational 
field into $D$ in its simplest form: we take its eigenvalues 
$\lambda_{n}$ as the dynamical variables of general relativity.  The 
main reason for which we think this is interesting is that these 
eigenvalues form an infinite family of {\em diffeomorphism invariant\ 
} observables.  Objects of this kind have long been sought for 
describing the geometry.  General relativity teaches us that 
fundamental physics is invariant under active diffeomorphisms.  
Physically, this means that there is no fixed nondynamical structure 
with respect to which location or motion could be defined 
\cite{berg,rel}.  Starting with Peter Bergamann's pioneering work 
\cite{berg}, a fully diffeomorphism invariant description of the 
geometry, free from the gauge redundancy of the usual formalism, has 
long been sought \cite{search}, but without much success so far.  Such 
a description would be precious, in particular, for quantum gravity 
\cite{diff}.  Now, the $\lambda_{n}$'s are precisely diffeomorphism 
invariant quantities. (In fact, the invariance is only under 
diffeomorphisms which preserve the spin structure; however, only large 
diffeomorphisms can change the spin structure.) 

As a first step towards using these ideas in classical and/or 
quantum theories, we derive an expression for the Poisson 
brackets of the Dirac eigenvalues.  We obtain this result by using the 
covariant formulation of the phase space of general relativity 
described in \cite{abhay} and by extending a technique developed in 
\cite{roberto}, where the eigenvalues of the three-dimensional Weyl 
operator, invariant under spatial diffeomorphism only, where analyzed.  
Surprisingly, we find that the Poisson brackets of the eigenvalues can 
be expressed explicitly in terms of the energy-momentum tensors of the 
corresponding Dirac eigenspinors.  These tensors form the Jacobian 
matrix of the change of coordinates between metric and eigenvalues.  
The brackets are quadratic in these tensors, with a kernel given by 
the propagator of the linearized Einstein equations.  The 
energy-momentum tensor of the Dirac eigenspinors provides therefore a 
key tool for analyzing the representation of spacetime geometry in 
terms of Dirac eigenvalues.

We also study a variant of the Chamseddine-Connes spectral action 
\cite{alain}.  In its simplest version, this action is a bit 
unrealistic because of a huge cosmological term.  This term is 
disturbing not only phenomenologically, but also because it implies 
that the small-curvature geometries for which the spectral action 
approximates the Einstein-Hilbert action are {\em not\/} solutions of 
the theory.  We study a version of the spectral action which 
eliminates the cosmological term.  We close by analyzing the equations 
of motion, derived from this action.  These are solved if the energy 
momenta of the high mass eigenspinors scale linearly with the mass.  
This scaling requirement approximates the vacuum Einstein equations.  
Thus, we obtain a representation of Einstein equations as a scaling 
law. Finally, we briefly describe the coupling of a Dirac fermion. 

Our results suggest that the Chamseddine-Connes gravitational theory 
can be viewed as a manageable theory, possibly with powerful 
applications to classical and quantum gravity.  The theory reproduces 
general relativity at low energies; it is formulated in terms of fully 
diffeomorphism invariant variables; and, of course, it prompts 
fascinating extensions of the very notion of geometry.  

A condensed version of some of the results presented here has appeared in
\cite{laro}.  An extension of these ideas to supergravity has been
considered in \cite{supergravity}.

\section{General Relativity in terms of Eigenvalues}

Consider Euclidean general relativity on a compact 4d (spin-) manifold 
$M$ without boundary.  For definitness, let us assume that $M$ has the 
topology $M=S^{3}\times S^{1}$ (the manifold that may be relevant in 
studying the finite temperature quantum properties of a compact 
Robertson-Walker universe).  The metric field is $g_{\mu\nu}(x)= 
E_{\mu}^{I}(x) E_{\nu\,I}(x)$, with $E_{\mu}^{I}(x)$ the tetrad 
fields.  Indices $\mu=1,\ldots,4$ are curved while $I=1,\ldots,4$ are 
internal euclidean, raised and lowered by the Kronecker metric 
$\delta_{IJ}$.  The spin connection $\omega_{\mu}^{I}{}_{J}$ is 
defined by $\partial_{[\mu}E_{\nu]}^{I} = \omega_{[\mu}^{I}{}_{J} 
E_{\nu]}^{J}$, where square brackets indicate anti-symmetrization.  
The dynamics is determined by the Einstein-Hilbert action
\begin{equation}
    S_{EH}[E]=\frac{1}{16\pi G}\	\int d^4 x\ \sqrt{g}\ R\ ,
	\label{EinstHilb}
\end{equation}
where $g$ and $R$ are the determinant and the Ricci
scalar of the metric respectively, $G$ is the Newton constant. 
We put the speed of light equal to one. The Planck constant does 
not enter our considerations, which are purely classical. 

In spite of a widespread contrary belief, the phase space is a 
spacetime covariant notion: it is the space of the solutions of 
the equations of motion modulo gauge transformations 
\cite{phase}.  Here, gauge transformations are 4d diffeomorphisms 
and local rotations of the tetrad field.  Thus, the phase space 
$\Gamma$ of general relativity is the space of the tetrad fields 
that solve Einstein equations, modulo internal rotations and 
diffeomorphisms.  Equivalently, it is the space of the Ricci flat 
``4-geometries''.  We denote the space of smooth tetrad 
fields as $\cal E$, the space of the tetrad fields that solve the 
Einstein equations as $\cal S$ (Solutions) and the space of the 
orbits of the gauge group in $\cal E$ as $\cal G$ (4-Geometries).  
The phase space $\Gamma$ is the space of the orbits of the 
diffeomorphism group in $\cal S$, or, equivalently, the subspace 
of $\cal G$ determined by the Einstein equations.  By 
definition, observables are functions on $\Gamma$ \cite{rel}.

We now define an infinite family of such observables. Consider the 
Hilbert space ${\cal H}$ of spinor fields $\psi$ on $M$. The scalar 
product is 
\begin{equation}
(\psi,\phi)=\int~d^{4}x\ \sqrt{g}\ ~\overline{\psi(x)}\phi(x),
\label{product}
\end{equation}
with bar indicating complex conjugation, and the scalar product in 
spinor space being the natural one in $C^{4}$.  With $\gamma^{I}$ 
being the (Euclidean) hermitian Dirac matrices, the curved Dirac 
operator is 
\begin{equation} 
D=\imath\,\gamma^{I}E^{\mu}_{I}\left(\partial_{\mu}+\omega_{\mu\,
JK}\,\gamma^{J}\gamma^{K}\right).
\end{equation}
The operator $D$ is a self-adjoint operator on ${\cal H}$  admitting a  
complete set of real eigenvalues $\lambda_{n}$ and ``eigenspinors'' 
$\psi_{n}$.  The manifold $M$ being compact, the spectrum is  discrete
\begin{equation}
D \psi_{n}=\lambda_{n}\ \psi_{n}~, \ \ \ \ \ \   
\end{equation}
The eigenvalues are labeled so that $\lambda_n \leq \lambda_{n+1}$, 
with repeated multiplicity.  Here $n$ is integer (positive and 
negative) and we choose $\lambda_{0}$ to be the positive eigenvalue 
closest to zero. For simplicity we assume that there are no zero modes. 
The eigenvalues have dimension of an inverse length. 

Notice that the Dirac operator depends on the gravitational field $E$; 
so do, of course, its eigenvalues as well.  We indicate explicitly 
this dependence by writing $D[E]$ and $\lambda_{n}[E]$. The latter 
defines a discrete family of real-valued functions on $\cal E$, 
$\lambda_{n}:E\longmapsto\lambda_{n}[E]$.  Equivalently, we have a function
$\lambda$ from $\cal E$ into the space of infinite sequences $R^\infty$
\begin{equation}
\begin{array}{cccc}
	\lambda: &  {\cal E} & \longrightarrow & R^\infty  \\
	 & 	E_{\mu}^{I}(x) & \longmapsto & \lambda_{n}[E].
\end{array}
	\label{lambda}
\end{equation}
The image $\lambda({\cal E})$ of $\cal E$ under this map is contained 
in the cone $\lambda_{n}\leq\lambda_{n+1}$ of $R^\infty$.  The 
functions $\lambda_{n}$ are invariant under 4d diffeomorphisms and 
under internal rotations of the tetrads.  Therefore they are gauge 
invariant and they are well defined functions on $\cal G$.  In 
particular, they are well defined on the phase space of general 
relativity $\Gamma$: thus, they are {\em observables} of pure general 
relativity.

Two metric fields with the same collection $\{\lambda_{n}\}$ are 
called ``isospectral''.  Isometric $E$ fields are isospectral, but the 
converse needs not be true \cite{drum,Gi}.  Therefore 
$\lambda$ might not be injective even if restricted to $\cal G$.  The 
$\lambda_{n}$'s might fail to coordinatize $\cal G$: they might fail to 
coordinatize $\Gamma$ as well, although whether or not this happens 
is not clear to us.  However, they presumably ``almost do it''.  
Following Connes \cite{alain}, it is tempting to consider the physical 
hypothesis that isospectral Ricci flat 4-geometries, are 
physical indistinguishable (``Spectral hypothesis'').

\section{The Poisson Brackets}

A simplectic structure on $\Gamma$ can be 
constructed in covariant form~\cite{abhay}.   A vector 
field $X$ on $\cal S$ can be written as a differential operator
\begin{equation}
X  =  \int d^{4}x \  X_{\mu}^{I}(x)[E]\ \ \frac{\delta}{\delta 
E_{\mu}^{I}(x)} 
\end{equation}
where $X_{\mu}^{I}(x)[E]$ is any solution of the Einstein equations 
for the tetrad field, {\it linearized\/} over the background $E$.  
A vector field $[X]$ on $\Gamma$ is given by an equivalence class of 
such vector fields $X$, modulo linearized gauge transformations of
$X_{\mu}^{I}(x)$.  A linearized gauge transformation 
is given in tetrad general relativity by 
\begin{eqnarray}
E &\longmapsto& E + \delta_{\vec v,\vec\rho}E \\
\delta_{\vec v,\vec\rho} E_{\mu}^{I}(x) &=& 
{\cal L}_{\vec v} E_{\mu}^{I}(x) + 
\rho^{a}(x)f_{a}{}^{I}_{K} E_{\mu}^{K}(x),
\end{eqnarray}
where $\vec v$ is a vector field (generating an infinitesimal 
diffeomorphism), $\vec\rho$ is an $SO(4)$ Lie algebra valued scalar 
field ($a=1,\ldots,6$), generating infinitesimal 4d rotations, ${\cal 
L}$ is the Lie derivative, and $f_{a}{}^{I}_{K}$ the generators of the 
vector representation $SO(4)$.  Two linearized field $X$ and $Y$ 
(around $E$) are gauge equivalent if
\begin{equation}
Y = X + \delta_{\vec v,\vec\rho}E,  
\end{equation}
for some couple $(\vec v,\vec\rho)$.

The simplectic two-form $\Omega$ of general relativity is given 
by~\cite{abhay}
\begin{equation}
\Omega(X,Y) = \frac{1}{32\pi G}\ \int_{\Sigma}d^{3}\sigma\  n_{\rho}\ 
(X_{\mu}^{I}\ \overleftarrow{\overrightarrow{\nabla}}{}_{\tau}\ 
Y_{\nu}^{J})\ \epsilon^{\tau}{}_{IJ\upsilon}\, 
\epsilon^{\upsilon\rho\mu\nu} 
\label{ome}
\end{equation}
where $(X_{\mu}^{I}\ \overleftarrow{\overrightarrow{\nabla}}{}_{\tau}\ 
Y_{\nu}^{J}) \equiv (X_{\mu}^{I}\ \nabla_{\tau}\ Y_{\nu}^{J} - 
Y_{\mu}^{I}\ \nabla_{\tau}\ X_{\nu}^{J})$.  From now on we put $32\pi 
G=1$.  Both sides of (\ref{ome}) are functions of $E$, namely scalar 
functions on ${\cal S}$; this $E$ is used to transform internal indices 
into spacetime indices.  Here $\Sigma:\sigma\longmapsto x(\sigma)$ is 
chosen to be a (compact non-contractible) three-dimensional 
surface, such that, topologically, $M=\Sigma\times S^{1}$ (so that it 
gives the single non trivial 3-cycle of $M$), but otherwise arbitrary, 
and $n_{\rho}$ its normal one-form.  

$\Omega$ is degenerate precisely in the gauge directions; 
\begin{equation}
\Omega(X,Y)[E]=0 \ \ \ \rm{iff} \ \ \ Y = \delta_{\vec 
v,\vec\rho}E, 
\end{equation}
thus it defines a non-degenerate {\it simplectic\/} two form on the space 
of the orbits, namely on $\Gamma$.   
The coefficients of $\Omega$ form can be written as
\begin{equation}
\Omega^{\mu\nu}_{IJ}(x,y) = \! \int_{\Sigma}\! d^{3}\sigma\ n_{\rho}\  
[\delta(x,x(\sigma)) \overleftarrow{\overrightarrow{\nabla}}_{\tau}  
\delta(y, x(\sigma)) ]\ \epsilon^{\tau}{}_{IJ\upsilon} 
\,\epsilon^{\upsilon\rho\mu\nu} .
\label{omega}
\end{equation} 
Because of the degeneracy, $\Omega$ has no inverse on $\cal S$. 
However, let us (arbitrarily) fix a gauge (choose a representative 
field $E$ for any four geometry, and, consequently, choose a field 
$X$ in any equivalence class $[X]$).   On the space of the 
gauge fixed fields, $\Omega$ is non degenerate and we can invert 
it. Let $P_{\mu\nu}^{IJ}(x,y)$ be the inverse of the simplectic form 
matrix on this subspace, namely
\begin{equation}
\int d^{4}y \int d^{4}z\ P_{\mu\nu}^{IJ}(x,y)\ 
\Omega^{\nu\rho}_{JK}(y,z) \ 
F_{\rho}^{K}(z) = \int d^{4}z \ \delta(x,z)\ 
\delta_{\mu}^{\rho}\ \delta^{I}_{K} \ F_{\rho}^{K}(z).
\end{equation}
for all solutions of the linearized Einstein equations $F$, 
satisfying the gauge condition chosen.  Integrating over the 
delta functions, and using (\ref{omega}), we have
\begin{equation}
\int_{\Sigma} d^{3}\sigma \ n_{\rho}\ [ 
P^{IJ}_{\mu\nu}(x,x(\sigma)) \overleftarrow{\overrightarrow{\nabla}}_{\rho} 
F^{K}_{\tau}(x(\sigma)) ]\ 
\epsilon^{\rho}{}_{JK\upsilon}\,\epsilon^{\upsilon\nu\tau\sigma} 
=  F_{\mu}^{I}(x). 
\label{p}
\end{equation}
This equation, where $F$ is any solution of the linearized equations, 
defines $P$, in the chosen gauge. 
Then, we can write the Poisson bracket between two 
functions $f, g$ on $\cal S$ as 
\begin{equation}
\{f,g\}= \int d^{4}x\int d^{4}y\ \ P_{\mu\nu}^{IJ}(x,y) \ 
\frac{\delta f}{\delta E_{\mu}^{I}(x)}\ 
\frac{\delta g}{\delta E_{\nu}^{I}(y)}.
\label{pp}
\end{equation}
If the functions $f$ and $g$ are gauge invariant, namely well 
defined on $\Gamma$, then the r.h.s\ of (\ref{pp}) is independent 
of the gauge chosen. 
But equation (\ref{p}) is precisely the definition of the 
propagator of the linearized Einstein equations over the background 
$E$, in the chosen gauge.   For instance, let us choose the surface 
$\Sigma$ as $x^{4}=0$ and fix the gauge with
\begin{equation}
X^{4}_{4}  =  1, \ \ 
X^{4}_{a} = 0, \ \ 
 X^{i}_{4} = 1, \ \  
 X^{i}_{a} = 0.
\end{equation}
where $a=1,2,3$ and $i=1,2,3$.  Then equation (\ref{p}) becomes
\begin{equation}
F^{i}_{a}(\vec x, t)=
\int d^3\vec y\ [P^{ib}_{aj}(\vec x, t; \vec y, 0) 
\overleftarrow{\overrightarrow{\nabla}}_{0} F^{j}_{b}(\vec y, 0)],
\end{equation}
where we have used the notation $\vec x = (x^1, x^2, x^3)$ and 
$t=x^4$, and the propagator can be easily recognized. 

Next, we need the Jacobian of the transformation from metric to eigenvalues.
The variation of $\lambda_{n}$ for a variation of $E$ can be computed 
using standard time independent quantum mechanics perturbation theory.
For a self-adjoint operator $D(v)$ depending on a parameter $v$ and whose 
eigenvalues $\lambda_{n}(v)$ are non-degenerate, we have
\begin{equation}
\frac {d\lambda_{n}(v)}{d v} = (\psi_{n}(v)| \left(\frac{d}{dv} 
D(v)\right)|\psi_{n}(v)).
\label{sedici}
\end{equation}
This equation is well known for its application in elementary 
quantum mechanics. It can be obtained by varying $v$ in the eigenvalue 
equation for $D(v)$, taking the scalar product with one of the 
eigenvectors, and noticing that the terms with the variation of the 
eigenvectors cancel.  We now apply this equation to our situation, 
assuming generic metrics with non-degenerate eigenvalues (we refer to 
\cite{diffeo} for the general situation).   We want to compute the 
variation of $\lambda_{n}[E]$ for a small variation of the tetrad 
field $E$.  Let  $\hat E_{\mu}^{I}(x)$ be an arbitrarily chosen tetrad
field and $v$ a real parameter, and consider the one parameter family 
of tetrad fields $E_{v}$ with components
\begin{equation}
	E_{v}{}_{\mu}^{I}(x) \equiv E_{\mu}^{I}(x)+v \hat E_{\mu}^{I}(x),
\end{equation}
Under the standard definition of functional derivative, the variation 
${\delta \lambda_{n}[E]}/{\delta E_{\mu}^{I}(x)}$ of the eigenvalues 
under a variation of the tetrad, is the distribution defined by 
\begin{equation}
\int d^{4}x \ \frac{\delta \lambda_{n}[E]}
{\delta E_{\mu}^{I}(x)} \ \hat E_{\mu}^{I}(x) =
\left.\frac{d \lambda_{n}[E_v]}{dv}\right|_{v=0}
\label{diciassette}
\end{equation}
Using (\ref{sedici}), we have 
\begin{equation}
\frac{d \lambda_{n}[E_v]}{dv}=
(\psi_{n}[E_v] | \frac{d D[E_v]}{dv}|\psi_{n}[E_v]).  
\end{equation} 
Explicitely 
\begin{equation}
\frac{d \lambda_{n}[E_{v}]}{dv}  = 
\int d^{4}y\ \sqrt{g[E_{v}]}\ \bar\psi_{n}[E_v]\ \frac{d D[E_v]}{dv}\ 
\psi_{n}[E_v].
\end{equation}
In $v=0$ we have
\begin{equation}
\left.\frac{d \lambda_{n}[E_{v}]}{dv}\right|_{v=0}  = 
\int d^{4}y\ \sqrt{g[E]}\ \bar\psi_{n}[E]\ 
\left.\frac{dD[E_v]}{dv}\right|_{v=0}\  \psi_{n}[E].
\end{equation}
Assuming the tetrad fields are suitably well behaved, 
we can rewrite this equation as 
\begin{eqnarray}
\left. \frac{d\lambda_{n}[E_{v}]}{dv}\right|_{v=0}  &=& 
\left. \frac{d}{dv}\right|_{v=0} \int d^{4}y\ \sqrt{g[E_{v}]}\ \bar\psi_{n}[E]\ 
D[E_v]\  \psi_{n}[E] 
\nonumber \\  
&& -   \int d^{4}y\ \left.\frac{d\sqrt{g[E_{v}]}}{dv}\right|_{v=0}\ 
\bar\psi_{n}[E]\  D[E]\  \psi_{n}[E] 
\nonumber \\  
& = & 
\left. \frac{d}{dv}\right|_{v=0} \int d^{4}y\ \sqrt{g[E_{v}]}\ \bar\psi_{n}[E]\ 
D[E_v]\  \psi_{n}[E] 
\nonumber \\  
&& -   \int d^{4}y\ \left.\frac{d\sqrt{g[E_{v}]}}{dv}\right|_{v=0}\ 
\bar\psi_{n}[E]\  \lambda_{n}[E]\  \psi_{n}[E] 
\nonumber \\  
& = & 
\left.\frac{d}{dv}\right|_{v=0} \int d^{4}y \sqrt{g[E_{v}]}\ 
(\bar\psi_{n} D[E_{v}] \psi_{n} - \lambda_{n} \bar\psi_{n}\psi_{n}). 
\label{diciotto}
\end{eqnarray}
The last formula gives the variation of the action of a spinor field 
with mass $\lambda_{n}$ under a variation of the metric, (computed for 
the $n$-th eigenspinor of $D[E]$).  But the variation of the action 
under a variation of the tetrad is a well known quantity in general 
relativity: it provides the general definition of the energy momentum tensor 
(here in tetrad notation) $T^{\mu}_{I}(x)$.  Indeed, the Dirac 
energy-momentum tensor (density) is defined in general by
\begin{equation}
T^{\mu}_{I}(x)\equiv\frac{\delta}{\delta E_{\mu}^{I}(x)}S_{\rm Dirac},
\label{deft}
\end{equation}
where $S_{\rm Dirac}=\int \sqrt{g}\ (\bar \psi D \psi - 
\lambda\bar\psi\psi)$ is the Dirac action of a spinor with 
``mass'' $\lambda$. (Since we have not put the Planck constant 
in the Dirac action, $\lambda$ has dimensions of an inverse length,
rather than a mass.) 
See for instance \cite{stanley}, where the explicit 
form of this tensor is also given. 
By denoting the energy momentum tensor of the 
eigenspinor $\psi_{n}$ as $T_{n}{}^{\mu}_{I}(x)$, we have obtained, 
from (\ref{diciassette}), (\ref{diciotto}) and (\ref{deft})
\begin{equation}
 \frac{\delta \lambda_{n}[E]}
{\delta E_{\mu}^{I}(x)} = T_{n}{}^{\mu}_{I}(x).
\label{t}
\end{equation}
This equation gives the variation of the eigenvalues 
$\lambda_{n}$ under a variation of the tetrad $E_{\mu}^{I}(x)$, 
namely the Jacobian matrix of the map $\lambda$.  The matrix 
elements of this Jacobian are given by the energy momentum tensor 
of the Dirac eigenspinors.  This fact suggests that we can study 
the map $\lambda$ locally in the space of the metrics, by 
studying the space of the eigenspinor's energy-momenta.  As far 
we know, little is known on the topology of the space of 
solutions of Euclidean Einstein's equations on a compact 
manifold.  A local analysis in $\Gamma$ would of course miss 
information on disconnected components of $\Gamma$. 

In order to avoid a possible confusion, we remark that 
the quantities $\lambda_{n}$ are invariant under 
diffeomorphisms, not under {\em arbitrary\/} 
changes of the metric or the tetrad fiels: the left hand side
of (\ref{t}) does not vanish in general.  Finally, notice that 
the above derivations would go through for several other 
operators, beside the Dirac operator. In \cite{moretti} a formula similar 
to (\ref{t}) has been derived for any second order elliptic selfadjoint 
operator. 

By combining (\ref{p},\ref{pp}) and (\ref{t}) we obtain our main result
(in this equation we restore physical units for completeness): 
\begin{equation}
\{\lambda_{n},\lambda_{m}\}= {\textstyle 32\pi G}
\int\!\! d^{4}x\!\!  
\int\!\! d^{4}y \ T_{[n}{}^{\mu}_{I}(x)\ P_{\mu \nu}^{IJ}(x, y) \ 
T_{m]}{}^{\nu}_{J}(y)
\label{main}
\end{equation}
which gives the Poisson bracket of two eigenvalues of the Dirac 
operator in terms of the energy-momentum tensor of the two 
corresponding eigenspinors and of the propagator of the linearized 
Einstein equations.  The right hand side does not depend on the gauge 
chosen for $P$.

Finally, if the transformation $\lambda$ between the ``coordinates'' 
$E_{\mu}^{I}(x)$ and the ``coordinates'' $\lambda_{n}$ is locally 
invertible on the phase space $\Gamma$, we can write the simplectic 
form directly in terms of the $\lambda_{n}$'s as
\begin{equation}
\Omega=\Omega_{mn}\ d\lambda_{n} \wedge d\lambda_{m},
\end{equation}
where a sum over indices is understood, and 
where $\Omega_{mn}$ is defined by
\begin{equation}
\Omega_{mn}\ T_{n}{}^{\mu}_{I}(x)\ T_{m}{}^{\nu}_{J}(y)= 
\Omega^{\mu\nu}_{IJ}(x,y).
\end{equation}
Indeed, let $d E_{\mu}^{I}(x)$ be a (basis) one-form on $\Gamma$, 
namely the infinitesimal difference between two solutions of Einstein 
equations, namely a solution of the Einstein equations linearized 
over $E$.  We have then
\begin{eqnarray} 
\Omega & = & \int d^{4}x \int d^{4}y \ 
\Omega^{\mu\nu}_{IJ}(x,y)\ d E_{\mu}^{I}(x) \wedge d E_{\nu}^{J}(y)
\nonumber \\
	 & = &  \int d^{4}x \int d^{4}y 
\Omega_{mn}\ T_{n}{}^{\mu}_{I}(x)\ d E_{\mu}^{I}(x) \wedge 
T_{m}{}^{\nu}_{J}(y)\ d E_{\nu}^{J}(y) 
\nonumber \\
	 & = & \Omega_{mn} \ 
d\lambda_{n} \wedge d\lambda_{m}.
\end{eqnarray}
An explicit evaluation of the matrix $\Omega_{nm}$ would be of great interest.

\section{Action and Field Equations}

We now turn to the gravitational spectral action \cite{alain}.  This 
action contains a cutoff parameter $l_0$ with units of a length, which 
determines the scale at which the gravitational theory defined departs 
from general relativity.  We may assume that $l_0$ is the Planck 
length $l_{0}\sim 10^{-33}cm$ (although we make no reference to 
quantum phenomena in the present context).  We use also 
$m_{0}=1/l_{0}$, which has the same dimension as $D$ and 
$\lambda_{n}$.  The action depends also on a dimensionless cutoff 
function $\chi(u)$, which vanishes for large $u$.  
The spectral action 
is then defined as
\begin{equation}
S_G[D] = \kappa \ Tr\left[\chi({l_{0}^{2}\,D^2})\right]
\label{action1}.
\end{equation}
$\kappa$ is a multiplicative constant to be chosen to recover the 
right dimensions of the action and the multiplicative overall factor 
in (\ref{EinstHilb}).  

The action (\ref{action1}) approximates the Einstein-Hilbert action 
with a large cosmological term for ``slowly varying'' metrics  with 
small curvature (with respect to the scale $l_{0}$). Indeed, 
the heat kernel expansion~\cite{alain,Gi}, allows to write 
(see \cite{laro} for a different derivation),
\begin{equation}\label{spac2}
 S_G(D) = (l_0)^{-4} f_0 \kappa
~\int_M \sqrt{g}\, d x  \ 
+\ (l_0)^{-2} f_2 \kappa  ~\int_M R\ \sqrt{g}\, d x \ 
+\ \dots~~ . 
\end{equation}
The functions $f_0$ and $f_2$ are defined by $f_0 = {1 \over 4 \pi^2} 
\int_0^\infty \chi(u) u d u$ and $f_2 = {1 \over 48 \pi^2} 
\int_0^\infty \chi(u) d u$, the integrals being of the order of unity 
for the choice of cutoff function made.  The other terms in 
(\ref{spac2}) are of higher order in $l_{0}$.

The expansion (\ref{spac2}) shows that the action (\ref{action1}) is 
dominated by the Einstein-Hilbert action with a Planck-scale
cosmological term.  The presence of this term is a problem for the 
physical interpretation of the theory because the solutions of the 
equations of motions have Planck-scale Ricci scalar, and therefore 
they are {\em all\/} out of the regime for which the approximation 
taken is valid!  However, the cosmological term can be canceled by 
replacing the function $\chi$ with $\widetilde{\chi}$,
\begin{equation}
\widetilde{\chi}(u) = \chi(u) - \epsilon^2 \chi(\epsilon u)~,
\end{equation}
with $\epsilon << 1$. Indeed, one finds 
$\widetilde{f}_0=0$~, $\widetilde{f}_2=(1-\epsilon)f_2$. 
The modified action becomes
\begin{equation}\label{spacmod}
\widetilde{S}_G(D) =  {\widetilde{f}_2 \kappa 
\over l_{0}^{2}} ~\int_M
R\ \sqrt{g}\, d x ~\ +\ \dots\ \ .
\end{equation}
We obtain the Einstein-Hilbert action (\ref{EinstHilb}) by fixing
\begin{equation}
	\kappa = \frac{l_{0}^{2}}{16 \pi G \widetilde{f}_{2}} ~. 
	\label{kappa}
\end{equation} 
If $l_{0}$ is the Planck length $\sqrt{\hbar G}$, then 
$\kappa=\frac{3}{2} h $, where $h$ is the Planck constant, up to terms 
of order $\epsilon$.  Low curvature geometries, for which 
the expansion (\ref{spac2}) holds {\em are now } solutions of the 
theory.  Thus we obtain a theory that genuinely approximates pure 
general relativity at scales large compared to $l_0$.

Let us now consider the equations of motion derived from this 
action. Following our philosophy, we want to regard the $\lambda_{n}$'s 
as the gravitational variables.  The action can easily be expressed in 
terms of these variables: 
\begin{equation}
 \widetilde S_{G}[\lambda] =  \kappa \sum_n\ 
 \widetilde\chi(l_{0}^{2}\lambda_{n}^{2}).
 \label{action}
\end{equation}
However, we cannot obtain (approximate) Einstein equations by simply 
varying (\ref{action}) with respect to the $\lambda_{n}$'s: we must 
minimize (\ref{action}) on the surface $\lambda({\cal E})$, not on the 
entire $R^{\infty}$.  In other words, the $\lambda_{n}$'s are not 
independent variables: there are relations among them, and these 
relations among them code the complexity of general relativity.  We 
shall comment on these relations at the end of the paper.  We can 
still obtain the equations of motion by varying $\widetilde{S}_{G}$ 
with respect to the tetrad field:
\begin{equation}\label{eqmot0}
0  =  \frac{\delta \widetilde{S}_{G}}{\delta E_{\mu}^{I}(x)} 
 =  \sum_n\ \frac{\partial \widetilde{S}_{G}}{\partial \lambda_{n}} \ 
\frac{\delta \lambda_{n}}{\delta E_{\mu}^{I}(x)}
 =  \sum_n\ 
 \frac{d\widetilde{\chi}(l_{0}^{2}\lambda_{n}^{2})}{d \lambda_{n}}\ 
T_{n}{}^{\mu}_{I}(x).
\end{equation}
Defining $f(u) =: \frac{d}{du}\widetilde{\chi}(u)$, (\ref{eqmot0}) 
becomes 
\begin{equation}\label{eqmot}
\sum_{n} f(l_{0}^{2}\lambda_{n}^{2}) \ \lambda_{n} \ T_{n}{}^{\mu}_{I}(x) = 0. 
\label{ee}
\end{equation}
These are the Einstein equations in the Dirac eigenvalues formalism.

The simplest choice for the cutoff function $\chi(u)$ is to take 
it to be smooth and monotonic on $R^{+}$ with
\begin{equation}
	\chi(u) = \left\{ 
	    \begin{array}{ll}
		     1 & \mbox{if $u < 1 - \delta$}  \\
		     0 &  \mbox{if $u> 1 + \delta$} 
	    \end{array}\right.
\end{equation}
where $\delta<<1$.  Namely $\chi(u)$ is the smoothed-out 
characteristic function of the $[0,1]$ interval.
With this choice, the action (\ref{action1}) is essentially 
(namely up to corrections of order $\delta$) simply ($\kappa$ times)  
the {\em number\ } of eigenvalues $\lambda_{n}$ with absolute value 
smaller that $m_{0}$! Then the function $f(u)$ 
vanishes everywhere except on two narrow peaks.  A negative one (width 
$2\delta$ and height $1/2\delta$) centered at one; and a positive one 
(width $2\delta/\epsilon$ and height $\epsilon^{3}/2\delta$) around 
the arbitrary large number $1/\epsilon =: s >>1$.  The first of these 
peaks gets contributions from $\lambda_{n}$'s such that $\lambda_{n} 
\sim m_{0}$, namely from Planck scale eigenvalues.  The second from 
ones such that $\lambda_{n} \sim s m_{0}$.  Equations (\ref{eqmot}) 
are solved if the contributions of the two peaks cancel.  This happens 
if below the Planck scale the energy momentum tensor scales as
\begin{equation}
\lambda_{n(m_{0})} \rho(1)\ T_{n(m_{0})}{}^{\mu}_{I}(x) = s^{-2} 
\lambda_{n(sm_{0})}\rho(s)\ T_{n(sm_{0})}{}^{\mu}_{I}(x),
\end{equation} 
Here, $\rho(1)$ and $\rho(s)$ are the densities of eigenvalues of 
$l^{2}_{0}D^2$ at the two peaks and the index $n(t)$ is defined by 
\begin{equation}
         l_{0}\lambda^{2}_{n(t)} = t.  	
\end{equation}
For large $n$ the growth of the eigenvalues of the Dirac operator
is given by the Weyl formula $\lambda_{n}\sim\sqrt{2\pi}V^{-1/4}
n^{1/4}$, where $V$ is the volume.  
Using this, one derives immediately the eigenvalue 
densities, and simple algebra yields
\begin{equation}
T_{n}{}^{\mu}_{I}(x) = \lambda_{n}\ l_{0}\ T_{0}{}^{\mu}_{I}(x)\ .
\label{scaling}
\end{equation}
for $n>>n(m_P)$, where $T_{0}{}^{\mu}_{I}(x)=T_{n(m_{0})}{}^{\mu}_{I}(x)$
is the energy momentum at the Planck scale.  
We have shown that {\it the dynamical equations for the 
geometry are solved if below the Planck length the energy-momentum of 
the eigenspinors scales as the eigenspinor's mass.}  In other 
words, we have expressed the Einstein equations as a scaling 
requirement on the energy-momenta of the very-high-frequency 
Dirac eigenspinors.

We add a few considerations that shed some light on this scaling 
requirement.  Notice that $T_{n}{}_{\mu}^{I}$ is formed by a term 
linear in the derivatives of the spinor field and a term independent 
from these.  The latter is a function of $(\psi, E, \partial_{\mu} 
E)$, quadratic in $\psi$.
\begin{equation}
T_{n}{}_{\mu}^{I}=\bar\psi_{n}\gamma^{I}
{\scriptstyle \overleftarrow{\overrightarrow{\partial}}}_{\mu}\psi_{n} +
S_{n}{}_{\mu}^{I}[\psi, E, \partial  E].
\end{equation}
If we expand the last term around a point of the manifold with local 
coordinates $x$, covariance and dimensional analysis require that
\begin{equation}
S_{n}{}_{\mu}^{I} = 
c_{0} \lambda_{n} E_{\mu}^{I}+c_{1}\ R_{\mu}^{I}+ c_{2}\  
R\,E_{\mu}^{I}+O\left( \frac{1}{\lambda_{n}}\right).
\end{equation}
for some fixed expansion coefficients $c_{0}, c_{1}$ and $c_{2}$.  
Here $R_{\mu}^{I}$ is the Ricci tensor.  \\
To be convinced that terms 
of this form do appear, consider the following.
\begin{eqnarray}
T_{n}{}_{\mu}^{I} &=& \bar\psi_{n}\gamma^{I}D_{\mu}\psi_{n} +\ldots  
\nonumber \\
& = & (\lambda_{n})^{-1}\ 
 \bar\psi_{n} 
\gamma^{I}\gamma^{\nu}D_{\mu}D_{\nu}\psi_{n}+\ldots 
\nonumber \\
& = &  (\lambda_{n})^{-1}\ 
\bar\psi_{n} \gamma^{I}\gamma^{\nu}[D_{\mu},D_{\nu}]\psi_{n}  +\ldots
\nonumber \\
& = & (\lambda_{n})^{-1}\ 
\bar\psi_{n} \gamma^{I}\gamma^{\nu}R_{\mu\nu}\psi_{n} +\ldots
 \nonumber \\
&=&   (\lambda_{n})^{-1}\ 
\bar\psi_{n} 
\gamma^{I}\gamma^{\nu}R_{\mu\nu}^{JK}\gamma_{J}\gamma_{K}\psi_{n} +\ldots
 \nonumber \\ 
& = &
Tr\ \  \gamma^{I}\gamma^{\nu}R_{\mu\nu}^{JK}\gamma_{J}\gamma_{K} +\ldots
 \nonumber \\ 
& = &
R^{I}_{\mu} + \ldots\  
\end{eqnarray}
For sufficiently high $n$, the eigenspinors are locally approximated by
plane waves in local cartesian coordinates.  For these functions, if we
double the mass the frequency doubles as well:   
if $\lambda_{m}=t\ \lambda_{n} $, then $\partial_{\mu}\psi_{m}=t\ 
\partial_{\mu}\psi_{n}$.  It follows that in general the energy momentum 
scales as
\begin{equation}
T_{n}{}_{\mu}^{I} = t 
\left[\bar\psi_{n}\gamma^{I}{\scriptstyle
\overleftarrow{\overrightarrow{\partial}}}_{\mu}\psi_{n} + c_{0} 
\lambda_{n}E_{\mu}^{I}\right] 
+ \left[c_{1}\ R_{\mu}^{I}+ c_{2}\ R\,E_{\mu}^{I}\right] + 
O\left(\frac{1}{\lambda_{n}}\right).
\end{equation}
For large $\lambda_{n}$ we can disregard the last term, and therefore 
(\ref{scaling}) requires that the second square bracket vanishes.  
Taking the trace we have $R=0$, using which we conclude $ R^{I}_{\mu} 
= 0 $, which are the vacuum Einstein equations.  
Thus, the equations of motion are
solved if  the  scaling requirement on the high mass eigenspinors' energy
momenta is  satisfied, and this requirement, in turn, yields vacuum 
Einstein equations at low energy scale. 

\section{Matter couplings}

The spinors $\psi_{n}$ that appear in the previous sections do not 
represent physical fermions.  They are mathematical quantities used to 
capture aspects of the pure gravitational field.  In particular, there 
is no sense in which they act back on the geometry.  In oder to 
describe the physical system formed by a (classical) fermion field, 
say with ``mass'' (inverse wavelenght) $m$, interacting with general 
relativity, namely an interacting Dirac-Einstein system, we have to 
introduce a (physical) spinor field $\psi(x)$.  The action 
that governs the dynamics of a fermion field and its interaction with 
the gravitational field is the Dirac action
\begin{equation}
	 S_{\rm Dirac}[\psi,E] 
	 =\int \ (\bar \psi D \psi - m \bar\psi\psi) \ \sqrt{g}\ d^4x
	 = (\psi [D-m] \psi)
	\label{dirac}
\end{equation}
Therefore the  Dirac-Einstein system is governed by the total action
\begin{equation}
	S[D,\psi] = \widetilde S_G[D] +  S_{\rm Dirac}[D,\psi] =
     \kappa\, Tr\left[\chi({l_{0}^{2}\,D^2})\right] + (\psi [D-m] \psi).
\end{equation}

The natural thing to do in the context of the present formalism is to 
expand $\psi$ in the basis formed by the $\psi_{n}$.  Namely to write
\begin{equation}
	\psi(x) = \sum_{n} \ a_{n}\ \psi_{n}(x)
\end{equation}
and to describe gravity in terms of the $\lambda_{n}$'s 
and the fermion in terms of its components $a_{n}$.  
The action becomes 
\begin{equation}
    S[\lambda_{n}, a_{n}] =  \sum_n\ \left[  \kappa \ 
 \widetilde\chi(l_{0}^{2}\lambda_{n}^{2}) + (\lambda_{n}-m) |a_{n}|^{2}
 \right].
\end{equation}
The equations of motion are 
\begin{eqnarray}
	\sum_{n}\left[ 2\kappa l_{0}^{2}\ f(l_{0}^{2}\lambda_{n}^{2}) \ 
	\lambda_{n} + |a_{n}|^{2}\right] \ T_{n}{}^{\mu}_{I}(x)
	 & = & 0,
	\label{eef}  \\
	(\lambda_{n}-m) \ a_{n} & = & 0
	\label{de}.
\end{eqnarray}
Eq. (\ref{eef}) corresponds to the Einstein equations with a source and
(\ref{de}) is the Dirac equation on a curved spacetime. Notice that 
the latter is algebraic, and it can be solved immediately. In order for a 
solution to exist there should exist an $\hat n$ such that
\begin{equation}
	\lambda_{\hat n}=m.
\end{equation}
The solution is 
\begin{eqnarray}
&& a_{n} = 0 \ \ \ \ \ \ \  \mbox{for all $n\ne \hat n$}, \nonumber \\
&& a_{\hat n} = a \ \ \ \ \ \ \ \mbox{an arbitrary constant}.
	\label{solution}
\end{eqnarray}
This is not surprising: a fermion field of mass $m$ on a geometry 
characterized by the Dirac eigenvalues $\{\lambda_{n}\}$ is given 
precisely by the eigenspinor $\psi_{\hat n}$ with eigenvalue equal to 
$m$.   Using the solution of the Dirac equation, (\ref{eef}) becomes
\begin{equation}
	\widetilde{f}_{2}^{-1}l_{0}^{4} \sum_{n} f(l_{0}^{2}\lambda_{n}^{2}) \ 
	\lambda_{n} \ T_{n}{}^{\mu}_{I}(x) = 8 \pi G\  |a|^{2} 
	T_{\hat n}{}^{\mu}_{I}(x) 
\end{equation}
where we have used the value of $\kappa$ (\ref{kappa}).
>From the results of the previous section, we recognize the left hand 
side as the Einstein tensor; the right hand side is the energy momentum 
tensor of the fermion.  

In the presence of matter, the scaling law (\ref{scaling}) is altered.  
Using again the Weyl formula, we obtain with simple algebra
\begin{equation}
T_{n}{}^{\mu}_{I}(x) = \lambda_{n}\ l_{0}\ [T_{0}{}^{\mu}_{I}(x)\,  
+ \alpha \ |a|^{2}\, T_{\hat n}{}^{\mu}_{I}(x)],
\label{scaling2}
\end{equation}
where 
\begin{equation}
	\alpha = \frac{16 \pi^{3} G\ \widetilde{f}_{2}l_{0}}{V}. 
\end{equation}
Equation (\ref{scaling2}) is the ``scaling law'' form of the Einstein 
equations, modified by the matter source term. 

The extension of the theory to other conventional matter couplings should 
not be difficult, but we do not pursue it here. 

\section{Summary and perspectives}

We have discussed the possibility of describing gravity by means of 
the Dirac operator eigenvalues.  This possibility has been opened by 
the recent work of Connes and Chamseddine.  We think that these new 
ideas might open a novel window over the physics of spacetime and find 
applications in classical and quantum gravitation.  The main obstacle 
for a full development of this approach is its natural euclidean 
character, due to the fact that on a non-compact lorentzian spacetime 
the Dirac operator will not have discrete spectrum (but see \cite{eli} 
for `lorentzian' attempts).  However, the present formalism might 
still find a natural application in quantum or in thermal quantum 
physics.

We have elucidated some aspects of the dynamical structure of the 
theory in the $\lambda_{n}$ variables by computing their Poisson 
algebra.  This is given in equation (\ref{main}).  Perhaps a quantum 
theory could be constructed in a diff-invariant manner by studying 
representations of this algebra, as suggested by 
Connes and Isham \cite{ca}.  At present, the Poisson algebra is 
not given in closed form, since the right hand side of equation 
(\ref{main}) is not expressed in terms of the $\lambda_{n}$ 
themselves.  This difficulty could be faced by expanding the energy 
momentum tensors in terms of the eigenspinors themselves, as suggested 
by Hawkins \cite{eli2}.

We have also studied the equations of motion of (a version) of the 
Cham\-seddi\-ne-Connes spectral action.  This action defines a theory 
that approximates general relativity at large scale, where it could be 
used as a tool in classical gravity.  It would be interesting to 
explore the modifications to general relativity that it yields at 
short scale.  We have given the expression for a fermion coupling in 
this formalism.  We have found a puzzling and intriguing way of 
expressing the Einstein equations as a scaling law for the energy 
momenta of the ultra-high-frequency eigenspinors.

The striking feature of the formalism discussed here is that the 
theory is formulated in terms of diffeomorphism invariant quantities.  
The $\lambda_{n}$'s are a family of diffeomorphism invariant 
observables in euclidean general relativity, which is presumably 
complete or ``almost complete'' (it could fail to distinguish possible 
isospectral and not isometric geometries).  It should be possible, at 
least in principle, to represent ``physical observations'' in pure 
gravity as a function of the $\lambda_{n}$'s alone.  Another 
remarkable aspect of the spectral action is that it introduces a 
physical cutoff and an elementary physical length without breaking 
diffeomorphism invariance.  The spectral action cuts off all high 
frequency modes, but it does so in a diffeomorphic invariant manner 
without introducing background structures.  Since the number of the 
remaining modes is determined by the ratio of the spacetime volume to 
the Planck scale, one may expect that a theory of this sort could have 
infrared but not ultraviolet divergences in the quantum regime.  The 
quantum theory based on the spectral action is therefore very much worth 
exploring, we think.

The key open problem, in our view, is to better understand the map 
$\lambda$ given in (\ref{lambda}) and its range; namely the 
constraints that a sequence of real numbers $\lambda_{n}$ must 
satisfy, if it represents the spectrum of the Dirac operator of some 
geometry.  This problem can be addressed locally (in phase space) by 
studying the {\it tangent\ } map to $\lambda$.  We have show that this 
tangent map is given explicitly by the eigenspinor's energy-momenta.  
One could begin to study $\lambda$ around simple geometries, such as a 
flat 4-torus.  On a more general ground, the constraints on the 
$\lambda_{n}$'s are presumably the core of the formulation of the 
gravitational theory that we have begun to explore here.  They should 
be contained in Connes' axioms for $D$ in the axiomatic definition of 
a spectral triple \cite{alain}.  The equations in these axioms capture 
the notion of Riemannian manifold algebraically, and they should code 
the constraints satisfied by the $\lambda_{n}$.  Finding the explicit 
connection between the formalism studied here and Connes axioms' 
equations would be of great interest.

\vskip 1cm

\noindent
{\bf Acknowledgments}

We thank R~DePietri, J~Fr\"ohlich, D~Kastler and especially A~Connes, 
for suggestions and conversations.  
GL tanks all members of DAMTP and in particular Prof. M. Green for the 
kind hospitality in Cambridge.
Work supported by the Italian 
MURST and by NSF grant PHY-95-15506.

\end{document}